\documentclass[manuscript,screen]{acmart}
\usepackage{caption}
\usepackage{subcaption}

\AtBeginDocument{%
  \providecommand\BibTeX{{%
    \normalfont B\kern-0.5em{\scshape i\kern-0.25em b}\kern-0.8em\TeX}}}

\setcopyright{acmcopyright}
\copyrightyear{2023}
\acmYear{2023}
\acmDOI{}

\acmConference[CHI '23 Workshop: Integrating AI in Human-Human Collaborative Ideation]{Make sure to enter the correct
  conference title from your rights confirmation emai}{April 28,
  2023}{Hamburg, Germany}
\acmBooktitle{CHI '23 Workshop: Integrating AI in Human-Human Collaborative Ideation} 
\begin{document}

%%
%% The "title" command has an optional parameter,
%% allowing the author to define a "short title" to be used in page headers.

\title[AI as mediator]{AI as mediator between composers, sound designers, and creative media producers}

\author{Sebastian L{\"o}bbers}
\affiliation{%
  \institution{Centre for Digital Music, Queen Mary University of London}
  \streetaddress{Mile End Road, E1 4NS}
  \city{London}
  \country{United Kingdom}}
\email{s.lobbers@qmul.ac.uk}

\author{Mathieu Barthet}
\affiliation{%
  \institution{Centre for Digital Music, Queen Mary University of London}
  \streetaddress{Mile End Road, E1 4NS}
  \city{London}
  \country{United Kingdom}}
\email{m.barthet@qmul.ac.uk}

\author{Gy{\"o}rgy Fazekas}
\affiliation{%
  \institution{Centre for Digital Music, Queen Mary University of London}
  \streetaddress{Mile End Road, E1 4NS}
  \city{London}
  \country{United Kingdom}}
\email{george.fazekas@qmul.ac.uk}

%%
%% By default, the full list of authors will be used in the page
%% headers. Often, this list is too long, and will overlap
%% other information printed in the page headers. This command allows
%% the author to define a more concise list
%% of authors' names for this purpose.
\renewcommand{\shortauthors}{L{\"o}bbers et al.}

%%
%% The abstract is a short summary of the work to be presented in the
%% article.
\begin{abstract}
  Musical professionals who produce material for non-musical stakeholders often face communication challenges in the early ideation stage. Expressing musical ideas can be difficult, especially when domain-specific vocabulary is lacking. This position paper proposes the use of artificial intelligence to facilitate communication between stakeholders and accelerate the consensus-building process. Rather than fully or partially automating the creative process, the aim is to give more time for creativity by reducing time spent on defining the expected outcome. To demonstrate this point, the paper discusses two application scenarios for interactive music systems that are based on the authors' research into gesture-to-sound mapping.

\end{abstract}

%%
%% The code below is generated by the tool at http://dl.acm.org/ccs.cfm.
%% Please copy and paste the code instead of the example below.
%%
% \begin{CCSXML}
% <ccs2012>
%  <concept>
%   <concept_id>10010520.10010553.10010562</concept_id>
%   <concept_desc>Computer systems organization~Embedded systems</concept_desc>
%   <concept_significance>500</concept_significance>
%  </concept>
%  <concept>
%   <concept_id>10010520.10010575.10010755</concept_id>
%   <concept_desc>Computer systems organization~Redundancy</concept_desc>
%   <concept_significance>300</concept_significance>
%  </concept>
%  <concept>
%   <concept_id>10010520.10010553.10010554</concept_id>
%   <concept_desc>Computer systems organization~Robotics</concept_desc>
%   <concept_significance>100</concept_significance>
%  </concept>
%  <concept>
%   <concept_id>10003033.10003083.10003095</concept_id>
%   <concept_desc>Networks~Network reliability</concept_desc>
%   <concept_significance>100</concept_significance>
%  </concept>
% </ccs2012>
% \end{CCSXML}

\ccsdesc[500]{Applied computing~Sound and music computing}
\ccsdesc[300]{Human-centered computing~User interface programming}
\ccsdesc[300]{Human-centered computing~Collaborative and social computing devices}

%%
%% Keywords. The author(s) should pick words that accurately describe
%% the work being presented. Separate the keywords with commas.
\keywords{gesture-to-sound mapping, AI mediated collaboration, human-human ideation}

%% A "teaser" image appears between the author and affiliation
%% information and the body of the document, and typically spans the
%% page.
% \begin{teaserfigure}
%   \includegraphics[width=\textwidth]{sampleteaser}
%   \caption{Seattle Mariners at Spring Training, 2010.}
%   \Description{Enjoying the baseball game from the third-base
%   seats. Ichiro Suzuki preparing to bat.}
%   \label{fig:teaser}
% \end{teaserfigure}

\received{22 February 2023}
% \received[revised]{12 March 2009}
% \received[accepted]{5 June 2009}

%%
%% This command processes the author and affiliation and title
%% information and builds the first part of the formatted document.
\maketitle

\section{Introduction}

%MB: modifications
Composers and sound designers often work with, or for, people who do not have expert musical knowledge. For example, a composer may work with a filmmaker or choreographer to create an original score serving a narrative, and a sound designer may be hired by a company to produce audio branding suiting their corporate identity. In both situations, the two sides are likely to propose and share ideas in an early exploratory stage to define more precisely what the sounds or music should express. A potential bottleneck lies in the imbalance in musical knowledge between the two parties. Individuals can have strong opinions and preferences for sound and music but when questioned about the underlying reasons, semantic descriptions differ between non-musicians and musicians \cite{yang2021examining}. Compared to non-experts, musicians tend to rely on technical terms instead of everyday language which can be hard to decode for a composer or sound designer. While technical language (e.g. ``a piece in the Lydian mode'') can express ideas to domain experts, it is opaque to other stakeholders. In some cases, non-verbal forms of communication like gestures or body movement might be preferred, e.g. when producers describe spatial audio~\cite{deacon2022s}.  In practice, these difficulties are overcome by sharing sound and music material that represents intended ideas and can be referenced in the ideation process. However, this introduces two new issues: (1) how is this material found? Stakeholders may use their own knowledge of existing musical compositions (``internal'' catalogue of music), but this becomes less viable when searching for specific sounds; (2) despite having a reference in mind, a composer or sound designer may create material that ultimately does not suit the assignment. How can ideas be quickly adapted for prototyping, e.g. fitting composed music to a film or choreography, to reduce ``wasted'' work? 
\\ \\ 
Recent advancements in artificial intelligence (AI) present new opportunities for music professionals, for example by generating audio from text-prompts~\cite{agostinelli2023musiclm}, creating guitar tablatures in a specific genre~\cite{sarmento2023gtr}, visualising sound intelligently~\cite{garber2021audiostellar} or retrieving audio from graphical sketches~\cite{engeln2021similarity} and gestures~\cite{zbyszynskiGestureTimbreSpaceMultidimensional}. However, AI research often focuses on the optimisation of quantifiable metrics like audio quality or classification accuracy, rather than exploring its role for application in the ``real world''. Reflecting on the authors' own work that uses AI for gesture-to-sound mappings, two potential applications are imagined to aid human-human music and sound ideation. The paper shortly describes these applications in the next section and discusses them in the context of the \textit{Integrating AI in Human-Human Collaborative Ideation} workshop in Section~\ref{sec:discussion}.

\section{Case studies}\label{sec:casestudies}
This section presents two interactive music systems, \textit{Liquid Dance} and \textit{SketchSynth}, and explores their capabilities for collaborative music and sound ideation in two hypothetical case studies that are based on the first author's experience as a professional composer and sound designer.  

\subsection{Liquid Dance}

\begin{figure}[h]
    \centering
    \begin{subfigure}[b]{0.45\textwidth}
        \centering
        \includegraphics[width=\textwidth]{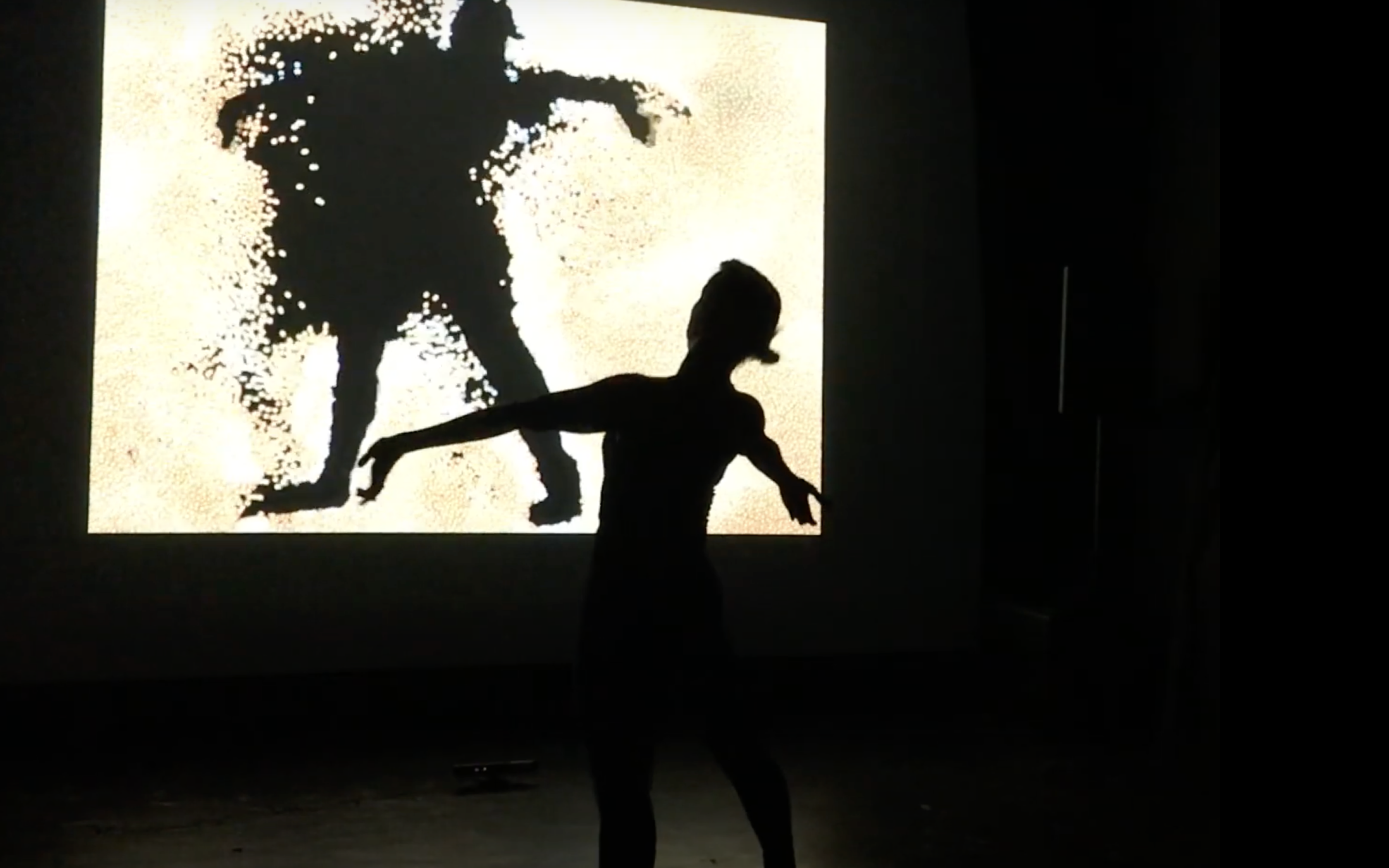}
        \caption{Video still}
        \label{fig:liquiddemonstration}
    \end{subfigure}
    \begin{subfigure}[b]{0.45\textwidth}
        \centering
        \includegraphics[width=\textwidth]{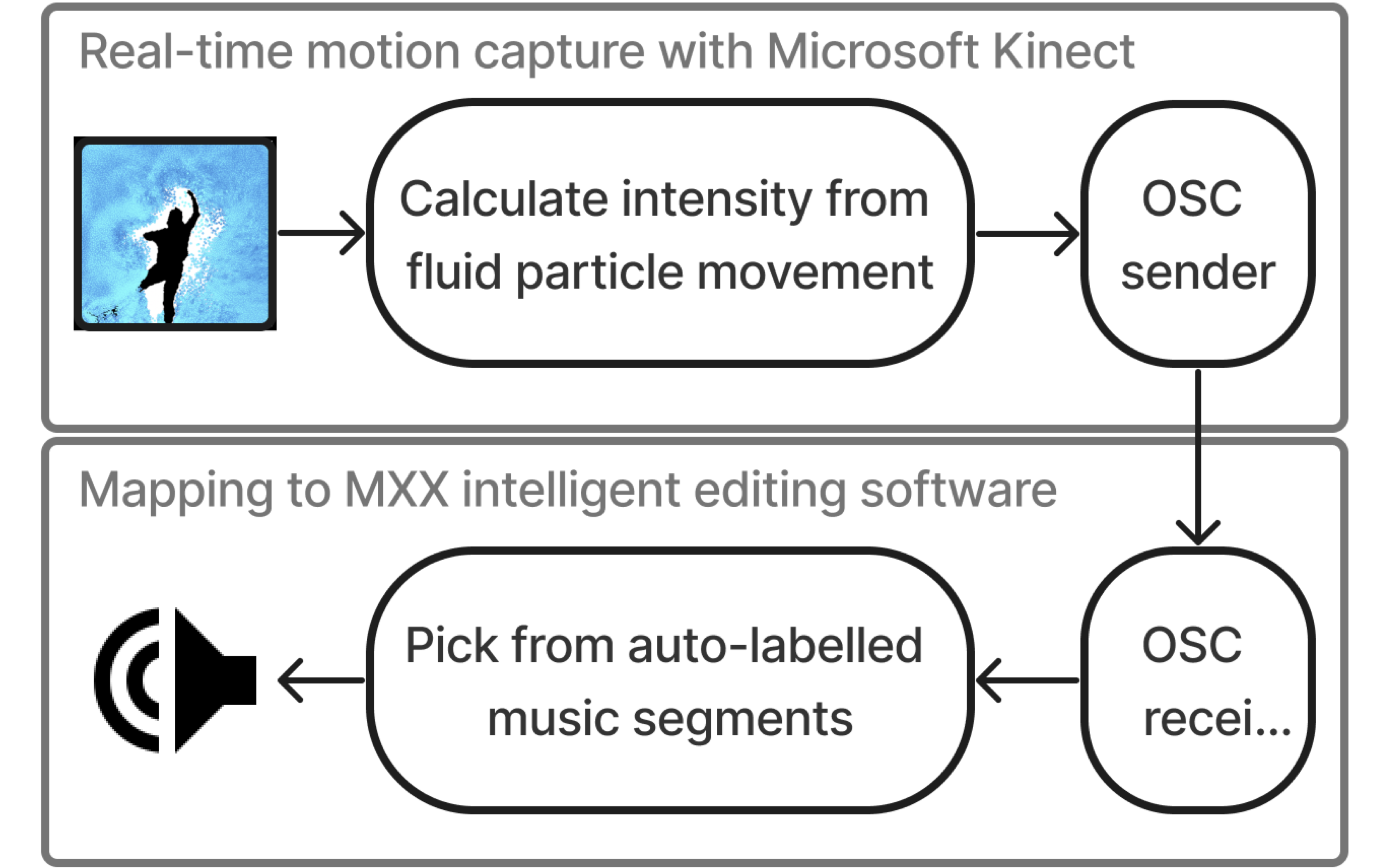}
        \caption{Flowchart}
        \label{fig:liquidflowchart}
    \end{subfigure}
    
    \caption{\textit{Liquid Dance} overview. Demonstration available at \url{https://youtu.be/fYInCkaOZO8}.}
    \label{fig:liquiddance}
\end{figure}

As illustrated in Figure~\ref{fig:liquiddance}, \textit{Liquid Dance} uses motion capture and a fluid particle simulation to quantify the intensity of dance movements and automatically assemble musical snippets creating a suitable score in real-time. It was developed in collaboration with a choreographer for the music AI startup MXX\footnote{\url{https://www.mxx.ai/}} to showcase their intelligent music editing software with an interactive performance. At a later stage, an unrelated composition project for dance sparked the idea of using \textit{Liquid Dance} for rapid testing of musical ideas for choreography. This is described in the following scenario:      
\\ \\
\textit{A choreographer and music composer have chosen several reference music pieces and are now interested in testing their compatibility with a choreographed dance performance. To achieve this, they plan on using a system that will enable them to dance alongside the music pieces, automatically selecting segments that best match the intensity of the dance movements. By using this method, they hope to gain a better understanding of how the selected music will work in tandem with the choreography.}

\subsection{SketchSynth}

\begin{figure}[h]
    \centering
    \begin{subfigure}[b]{0.45\textwidth}
        \centering
        \includegraphics[width=\textwidth]{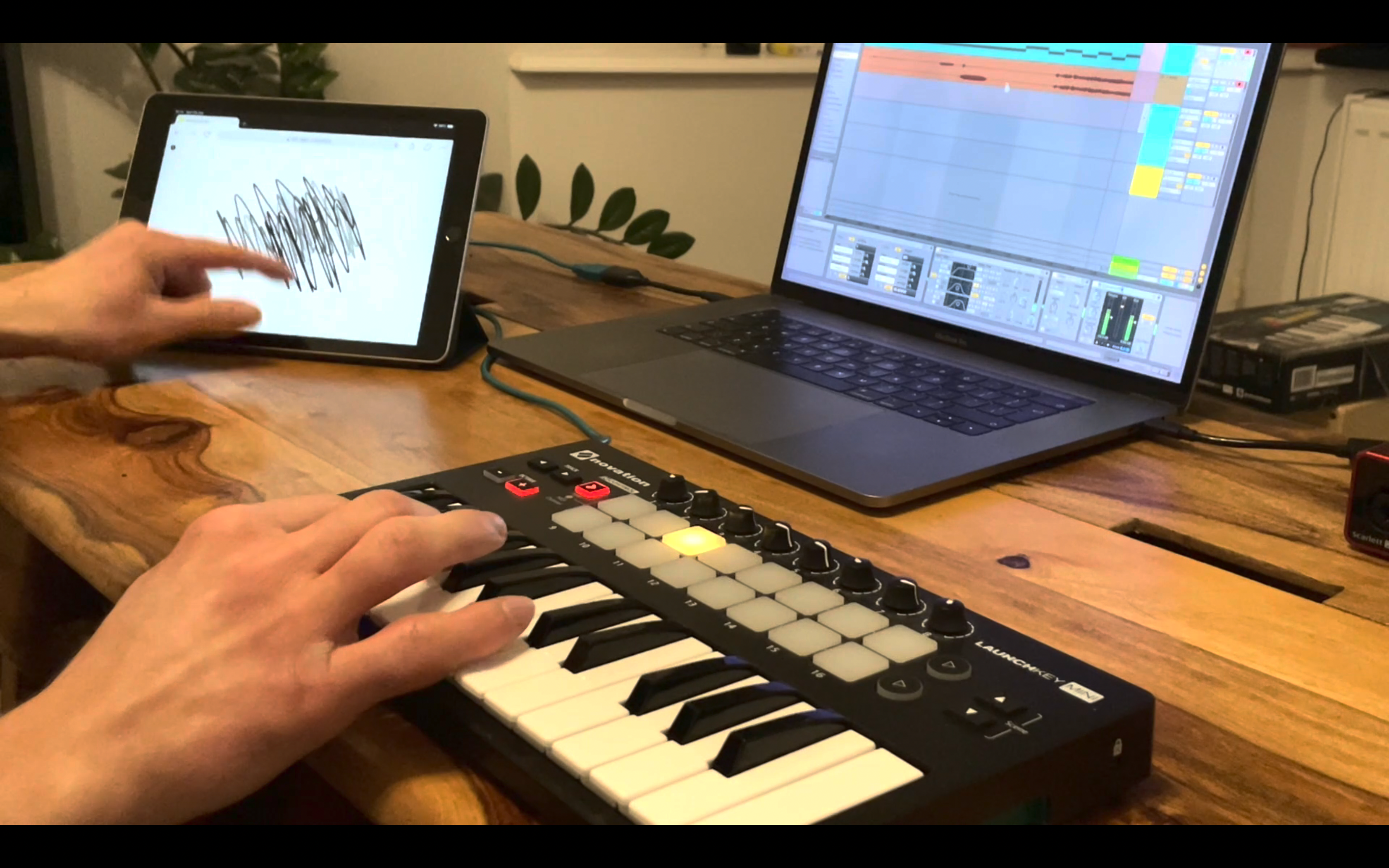}
        \caption{Video still}
        \label{fig:sketchdemonstration}
    \end{subfigure}
    \begin{subfigure}[b]{0.45\textwidth}
        \centering
        \includegraphics[width=\textwidth]{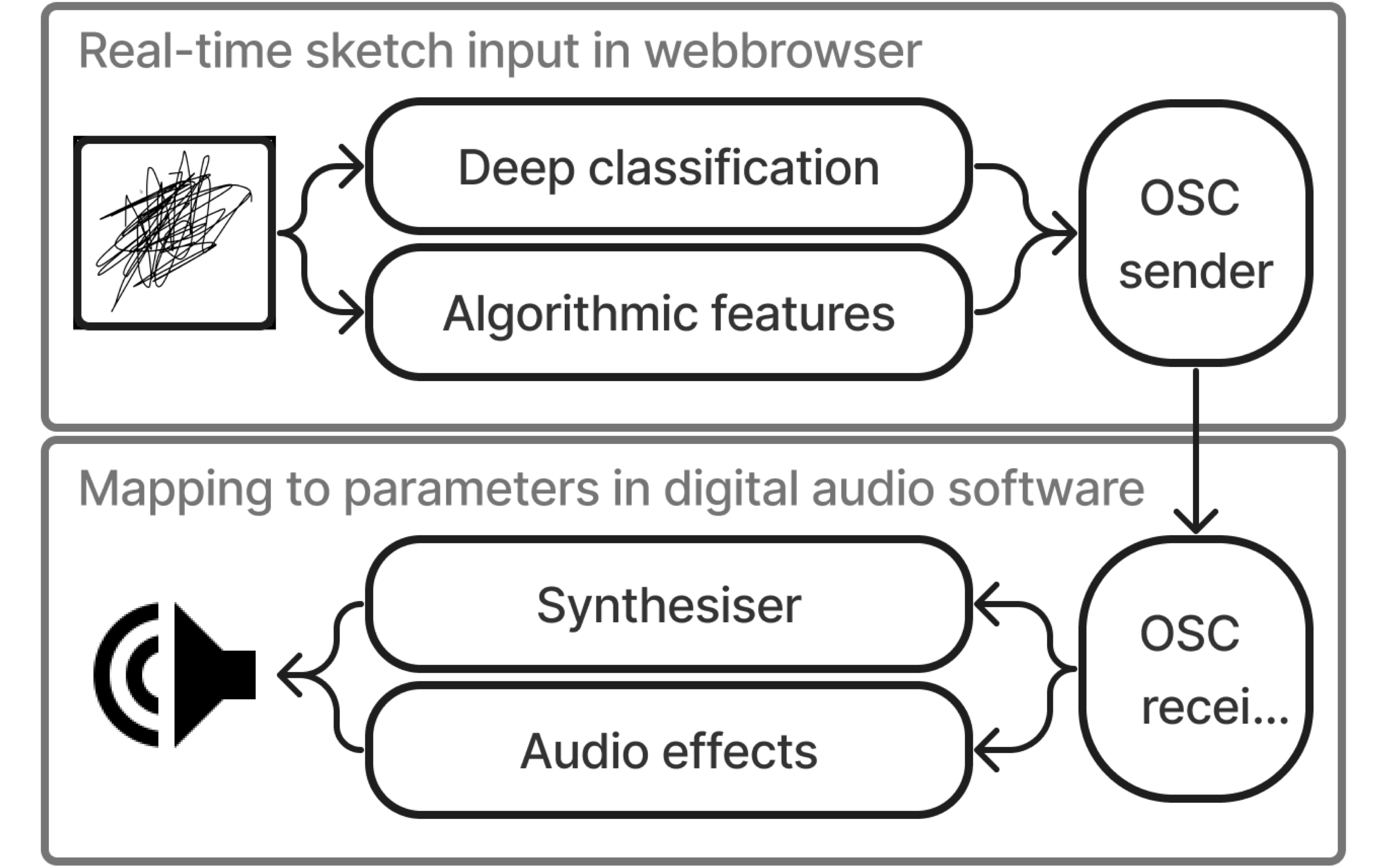}
        \caption{Flowchart}
        \label{fig:sketchflowchart}
    \end{subfigure}
    
    \caption{\textit{SketchSynth} overview. Demonstration available at \url{https://youtu.be/arSFt3iBAUM}. }
    \label{fig:sketchsynth}
\end{figure}
\textit{SketchSynth} is based on the authors' research into cross-modal perception of musical timbre~\cite{lobbers2021sketching, lobbers2022seeing, lobbers2023sketchsynth}. It employs AI and algorithmic methods to predict sound from a graphical sketch input as illustrated in Figure~\ref{fig:sketchsynth}. The following scenario describes a potential application in industrial sound design: 
\\ \\
\textit{A sound designer and product manager are looking for suitable sounds to incorporate into their product. To facilitate this process, they plan on using \textit{SketchSynth}, which enables non-experts to express their ideas and explore the sonic space through an intuitive mode of interaction. Both parties can literally draw out ideas until reaching a consensus from which the sound designer can elaborate in greater detail.}

\section{Discussion and conclusion}\label{sec:discussion}

In both case studies AI acts as a translator between non-verbal expression and sound or music creation. While an expert human actor is likely capable of creating suitable content from these expressions, a machine can be trained to perform this task much faster in real-time. This gives stakeholders the opportunity to test initial ideas much faster, focus on their interpersonal relationships and leave more time for the creative process that is arguably done best by a human professional. In this context, AI is seen strictly as a tool that is expected to reliably perform a complex but well-defined task. \textit{Liquid Dance} uses a robust, but very specific setup that might only be useful for a narrow type of projects; \textit{SketchSynth} tries to find universal sketch-to-sound mappings that might be too broad to communicate sound in sufficient detail. As user needs are likely to change between or even during projects, the scope of AI tools might have to change with them. This raises the question of not only how AI is used during human-human ideation, but also how it is trained and set up. One approach is to choose a model that was trained on a diverse dataset to solve a large number of projects. However, this might hand over too much agency leading to unexpected behaviour not suitable for the task at hand. Therefore, a user might want to adjust a system in an initial ``briefing'' stage to prepare for the upcoming task. Using the case studies in this paper as an example: for \textit{SketchSynth} this could mean confining the space of possible sounds, e.g. to mechanic sounds for an electronic car project or digital blips for a mobile game; or for \textit{Liquid Dance}, optimising the interface for a specific dance style, e.g. ballet or contemporary, and conditioning the AI-driven music segmentation on a specific genre like classical or electronic music. Ultimately, it remains for the user to define how much agency they want their system to display during the ideation process. In conclusion, recent AI-driven music research can find valuable application in aiding human-human ideation for example as a translator for non-verbal descriptions of sound and music. However, AI tools may have to be adjustable to user needs and project requirements to avoid unpredictable or inadequate behaviour.

%The bibliography is included in your source document with these two commands, placed just before the \verb|\end{document}| command:
%\begin{verbatim}
%\bibliographystyle{ACM-Reference-Format}
%  \bibliography{bibfile}
%\end{verbatim}
%where ``\verb|bibfile|'' is the name, without the ``\verb|.bib|''
%suffix, of the \BibTeX\ file.

\section{Acknowledgments}
EPSRC and AHRC Centre for Doctoral Training in Media and Arts Technology (EP/L01632X/1).
\bibliographystyle{ACM-Reference-Format}
\bibliography{chi2023}

%%% -*-BibTeX-*-
%%% Do NOT edit. File created by BibTeX with style
%%% ACM-Reference-Format-Journals [18-Jan-2012].

\begin{thebibliography}{10}

%%% ====================================================================
%%% NOTE TO THE USER: you can override these defaults by providing
%%% customized versions of any of these macros before the \bibliography
%%% command.  Each of them MUST provide its own final punctuation,
%%% except for \shownote{}, \showDOI{}, and \showURL{}.  The latter two
%%% do not use final punctuation, in order to avoid confusing it with
%%% the Web address.
%%%
%%% To suppress output of a particular field, define its macro to expand
%%% to an empty string, or better, \unskip, like this:
%%%
%%% \newcommand{\showDOI}[1]{\unskip}   % LaTeX syntax
%%%
%%% \def \showDOI #1{\unskip}           % plain TeX syntax
%%%
%%% ====================================================================

\ifx \showCODEN    \undefined \def \showCODEN     #1{\unskip}     \fi
\ifx \showDOI      \undefined \def \showDOI       #1{#1}\fi
\ifx \showISBNx    \undefined \def \showISBNx     #1{\unskip}     \fi
\ifx \showISBNxiii \undefined \def \showISBNxiii  #1{\unskip}     \fi
\ifx \showISSN     \undefined \def \showISSN      #1{\unskip}     \fi
\ifx \showLCCN     \undefined \def \showLCCN      #1{\unskip}     \fi
\ifx \shownote     \undefined \def \shownote      #1{#1}          \fi
\ifx \showarticletitle \undefined \def \showarticletitle #1{#1}   \fi
\ifx \showURL      \undefined \def \showURL       {\relax}        \fi
% The following commands are used for tagged output and should be
% invisible to TeX
\providecommand\bibfield[2]{#2}
\providecommand\bibinfo[2]{#2}
\providecommand\natexlab[1]{#1}
\providecommand\showeprint[2][]{arXiv:#2}

\bibitem[Agostinelli et~al\mbox{.}(2023)]%
        {agostinelli2023musiclm}
\bibfield{author}{\bibinfo{person}{Andrea Agostinelli}, \bibinfo{person}{Timo~I
  Denk}, \bibinfo{person}{Zal{\'a}n Borsos}, \bibinfo{person}{Jesse Engel},
  \bibinfo{person}{Mauro Verzetti}, \bibinfo{person}{Antoine Caillon},
  \bibinfo{person}{Qingqing Huang}, \bibinfo{person}{Aren Jansen},
  \bibinfo{person}{Adam Roberts}, \bibinfo{person}{Marco Tagliasacchi},
  {et~al\mbox{.}}} \bibinfo{year}{2023}\natexlab{}.
\newblock \showarticletitle{MusicLM: Generating Music From Text}.
\newblock \bibinfo{journal}{\emph{arXiv preprint arXiv:2301.11325}}
  (\bibinfo{year}{2023}).
\newblock


\bibitem[Deacon et~al\mbox{.}(2022)]%
        {deacon2022s}
\bibfield{author}{\bibinfo{person}{Thomas Deacon}, \bibinfo{person}{Patrick
  Healey}, {and} \bibinfo{person}{Mathieu Barthet}.}
  \bibinfo{year}{2022}\natexlab{}.
\newblock \showarticletitle{“It’s cleaner, definitely”: Collaborative
  Process in Audio Production}.
\newblock \bibinfo{journal}{\emph{Computer Supported Cooperative Work (CSCW)}}
  (\bibinfo{year}{2022}), \bibinfo{pages}{1--31}.
\newblock


\bibitem[Engeln et~al\mbox{.}(2021)]%
        {engeln2021similarity}
\bibfield{author}{\bibinfo{person}{Lars Engeln}, \bibinfo{person}{Nhat~Long
  Le}, \bibinfo{person}{Matthew McGinity}, {and} \bibinfo{person}{Rainer
  Groh}.} \bibinfo{year}{2021}\natexlab{}.
\newblock \showarticletitle{Similarity Analysis of Visual Sketch-based Search
  for Sounds}.
\newblock In \bibinfo{booktitle}{\emph{Proceedings of Audio Mostly 2021}}.
  \bibinfo{publisher}{Association for Computing Machinery},
  \bibinfo{address}{Trento, Italy}, \bibinfo{pages}{101--108}.
\newblock
\urldef\tempurl%
\url{https://doi.org/10.1145/3478384.3478423}
\showDOI{\tempurl}


\bibitem[Garber et~al\mbox{.}(2021)]%
        {garber2021audiostellar}
\bibfield{author}{\bibinfo{person}{Leandro Garber}, \bibinfo{person}{T
  Ciccola}, {and} \bibinfo{person}{JC Amusategui}.}
  \bibinfo{year}{2021}\natexlab{}.
\newblock \showarticletitle{AudioStellar, an open source corpus-based musical
  instrument for latent sound structure discovery and sonic experimentation}.
  In \bibinfo{booktitle}{\emph{Proceedings of International Computer Music
  Conference}}. \bibinfo{publisher}{Michigan Publishing Services},
  \bibinfo{address}{Santiago, Chile}, \bibinfo{pages}{62--67}.
\newblock
\urldef\tempurl%
\url{https://hdl.handle.net/2027/fulcrum.t435gg568}
\showURL{%
\tempurl}


\bibitem[L{\"o}bbers et~al\mbox{.}(2021)]%
        {lobbers2021sketching}
\bibfield{author}{\bibinfo{person}{Sebastian L{\"o}bbers},
  \bibinfo{person}{Mathieu Barthet}, {and} \bibinfo{person}{Gy{\"o}rgy
  Fazekas}.} \bibinfo{year}{2021}\natexlab{}.
\newblock \showarticletitle{Sketching sounds: an exploratory study on
  sound-shape associations}. In \bibinfo{booktitle}{\emph{Proceedings of
  International Computer Music Conference}}. \bibinfo{publisher}{Michigan
  Publishing Services}, \bibinfo{address}{Santiago, Chile},
  \bibinfo{pages}{299--304}.
\newblock
\urldef\tempurl%
\url{https://hdl.handle.net/2027/fulcrum.t435gg568}
\showURL{%
\tempurl}


\bibitem[L{\"o}bbers and Fazekas(2022)]%
        {lobbers2022seeing}
\bibfield{author}{\bibinfo{person}{Sebastian L{\"o}bbers} {and}
  \bibinfo{person}{Gy{\"o}rgy Fazekas}.} \bibinfo{year}{2022}\natexlab{}.
\newblock \showarticletitle{Seeing Sounds, Hearing Shapes: a gamified study to
  evaluate sound-sketches}. In \bibinfo{booktitle}{\emph{Proceedings
  International Computer Music Conference}}. \bibinfo{publisher}{Michigan
  Publishing Services}, \bibinfo{address}{Limerick, Ireland},
  \bibinfo{pages}{174--179}.
\newblock
\urldef\tempurl%
\url{https://hdl.handle.net/2027/fulcrum.nk322g689}
\showURL{%
\tempurl}


\bibitem[L{\"o}bbers and Fazekas(2023)]%
        {lobbers2023sketchsynth}
\bibfield{author}{\bibinfo{person}{Sebastian L{\"o}bbers} {and}
  \bibinfo{person}{Gy{\"o}rgy Fazekas}.} \bibinfo{year}{2023}\natexlab{}.
\newblock \showarticletitle{SketchSynth: cross-modal control of sound
  synthesis}.
\newblock \bibinfo{journal}{\emph{EvoMUSART preprint}} (\bibinfo{year}{2023}).
\newblock
\urldef\tempurl%
\url{https://sebastianlobbers.com/static/76293e97d37a3d69941f2f1b047cec7e/SketchSynth_cross-modal_control_of_sound_synthesis.pdf}
\showURL{%
\tempurl}


\bibitem[Sarmento et~al\mbox{.}(2023)]%
        {sarmento2023gtr}
\bibfield{author}{\bibinfo{person}{Pedro Sarmento}, \bibinfo{person}{Adarsh
  Kumar}, \bibinfo{person}{Yu-Hua Chen}, \bibinfo{person}{CJ Carr},
  \bibinfo{person}{Zack Zukowski}, {and} \bibinfo{person}{Mathieu Barthet}.}
  \bibinfo{year}{2023}\natexlab{}.
\newblock \showarticletitle{GTR-CTRL: Instrument and Genre Conditioning for
  Guitar-Focused Music Generation with Transformers}.
\newblock \bibinfo{journal}{\emph{arXiv preprint arXiv:2302.05393}}
  (\bibinfo{year}{2023}).
\newblock


\bibitem[Yang et~al\mbox{.}(2021)]%
        {yang2021examining}
\bibfield{author}{\bibinfo{person}{Simin Yang},
  \bibinfo{person}{Courtney~Nicole Reed}, \bibinfo{person}{Elaine Chew}, {and}
  \bibinfo{person}{Mathieu Barthet}.} \bibinfo{year}{2021}\natexlab{}.
\newblock \showarticletitle{Examining emotion perception agreement in live
  music performance}.
\newblock \bibinfo{journal}{\emph{IEEE transactions on affective computing}}
  (\bibinfo{year}{2021}).
\newblock


\bibitem[Zbyszy{\'n}ski et~al\mbox{.}(2021)]%
        {zbyszynskiGestureTimbreSpaceMultidimensional}
\bibfield{author}{\bibinfo{person}{Michael Zbyszy{\'n}ski},
  \bibinfo{person}{Balandino Di~Donato}, \bibinfo{person}{Federico~Ghelli
  Visi}, {and} \bibinfo{person}{Atau Tanaka}.} \bibinfo{year}{2021}\natexlab{}.
\newblock \showarticletitle{Gesture-{{Timbre Space}}: {{Multidimensional
  Feature Mapping Using Machine Learning}}}. In
  \bibinfo{booktitle}{\emph{Proceedings of International Symposium on Computer
  Music Multidisciplinary Research}}. \bibinfo{publisher}{Springer},
  \bibinfo{address}{Tokyo, Japan}, \bibinfo{pages}{600--622}.
\newblock
\urldef\tempurl%
\url{https://doi.org/10.1007/978-3-030-70210-6\_39}
\showDOI{\tempurl}


\end{thebibliography}

\end{document}